\documentclass[twocolumn]{aastex62}

\newcommand{\msol}{\,\mbox{M$_{\odot}$}}
\newcommand{\Ni}{\mbox{$^{56}$Ni}}

\newcommand{\dmr}{\mbox{$\Delta m_{15}(r) = \,$}}
\newcommand{\gmax}{\mbox{$g$-band maximum}}
\newcommand{\PS}{\protect \hbox {Pan-STARRS1}}
\newcommand{\tardis}{\protect \hbox {{\sc tardis}}}

\newcommand{\kms}{\mbox{$\rm{km}\,s^{-1}$}}

\def\lesssim{\mathrel{\hbox{\rlap{\hbox{\lower4pt\hbox{$\sim$}}}\hbox{$<$}}}}
\def\gtrsim{\mathrel{\hbox{\rlap{\hbox{\lower4pt\hbox{$\sim$}}}\hbox{$>$}}}}

\newcommand{\grizy}{\ensuremath{grizy_{\rm P1}}}

\graphicspath{{./}{figures/}}

\submitjournal{ApJ}

\shorttitle{The faint Iax supernova 2019gsc}
\shortauthors{Srivastav et al.}

\begin{document}

\title{The lowest of the low: discovery of SN 2019gsc and the nature of faint Iax supernovae}

\author{Shubham Srivastav}
\affiliation{Astrophysics Research Centre, School of Mathematics and Physics, Queen’s University Belfast, BT7 1NN, UK}

\author{Stephen J. Smartt}
\affiliation{Astrophysics Research Centre, School of Mathematics and Physics, Queen’s University Belfast, BT7 1NN, UK}

\author{Giorgos Leloudas}
\affiliation{DTU Space, National Space Institute, Technical University of Denmark, DK-2800 Kongens Lyngby, Denmark}

\author{Mark E. Huber}
\affiliation{Institute of Astronomy, University of Hawaii, 2680 Woodlawn Drive, Honolulu, HI 96822, USA}

\author{Ken Chambers}
\affiliation{Institute of Astronomy, University of Hawaii, 2680 Woodlawn Drive, Honolulu, HI 96822, USA}

\author{Daniele B. Malesani}
\affiliation{DTU Space, National Space Institute, Technical University of Denmark, DK-2800 Kongens Lyngby, Denmark}

\author{Jens Hjorth}
\affiliation{DARK, Niels Bohr Institute, University of Copenhagen, Lyngbyvej 2, DK-2100 Copenhagen \O, Denmark}

\author{James H. Gillanders}
\affiliation{Astrophysics Research Centre, School of Mathematics and Physics, Queen’s University Belfast, BT7 1NN, UK}

\author{A. Schultz}
\affiliation{Institute of Astronomy, University of Hawaii, 2680 Woodlawn Drive, Honolulu, HI 96822, USA}

\author{Stuart A. Sim}
\affiliation{Astrophysics Research Centre, School of Mathematics and Physics, Queen’s University Belfast, BT7 1NN, UK}

\author{Katie Auchettl}
\affiliation{DARK, Niels Bohr Institute, University of Copenhagen, Lyngbyvej 2, DK-2100 Copenhagen \O, Denmark}
\affiliation{Department of Astronomy and Astrophysics, University of California, Santa Cruz, CA 95064, USA}
\affiliation{School of Physics, The University of Melbourne, Parkville, VIC 3010, Australia}

\author{Johan P. U. Fynbo}
\affiliation{Cosmic Dawn Center (DAWN)}
\affiliation{Niels Bohr Institute, University of Copenhagen, Lyngbyvej 2, DK-2100 Copenhagen \O, Denmark}

\author{Christa Gall}
\affiliation{DARK, Niels Bohr Institute, University of Copenhagen, Lyngbyvej 2, DK-2100 Copenhagen \O, Denmark}

\author{Owen R. McBrien}
\affiliation{Astrophysics Research Centre, School of Mathematics and Physics, Queen’s University Belfast, BT7 1NN, UK}

\author{Armin Rest}
\affiliation{Space Telescope Science Institute, 3700 San Martin Drive, Baltimore, MD 21218, USA}
\affiliation{Department of Physics and Astronomy, Johns Hopkins University, Baltimore, MD 21218, USA}

\author{Ken W. Smith}
\affiliation{Astrophysics Research Centre, School of Mathematics and Physics, Queen’s University Belfast, BT7 1NN, UK}

\author{Radoslaw Wojtak}
\affiliation{DARK, Niels Bohr Institute, University of Copenhagen, Lyngbyvej 2, DK-2100 Copenhagen \O, Denmark}

\author{David R. Young}
\affiliation{Astrophysics Research Centre, School of Mathematics and Physics, Queen’s University Belfast, BT7 1NN, UK}

\correspondingauthor{s.srivastav@qub.ac.uk}





\begin{abstract}

We present the discovery and optical follow-up of the faintest supernova-like transient 
known. The event (SN 2019gsc) was discovered in a star-forming host at 53\,Mpc 
by ATLAS. A detailed multi-colour light curve was gathered with Pan-STARRS1 and follow-up spectroscopy was obtained with the NOT and Gemini-North. 
The spectra near maximum light show narrow features at low velocities of 3000 to 4000 km s$^{-1}$, 
similar to the extremely low luminosity SNe 2010ae and 2008ha, and the light curve 
displays a similar fast decline (\dmr $0.91 \pm 0.10$ mag).
SNe 2010ae and 2008ha have been classified as type Iax supernovae, and together the three either make up a distinct 
physical class of their own or are at the extreme low luminosity end of this diverse 
supernova population. 
The bolometric light curve is consistent with a low kinetic energy of 
explosion ($E_{\rm k} \sim 10^{49}$ erg s$^{-1}$), a 
modest ejected mass ($M_{\rm ej} \sim 0.2$ \msol) and 
radioactive powering by \Ni\ ($M_{\rm Ni} \sim 2 \times 10^{-3}$ \msol). The 
spectra are quite well reproduced with radiative transfer models (TARDIS) and a 
composition dominated by carbon, oxygen, magnesium, silicon and sulphur. 
Remarkably, all three of these extreme Iax events are in similar low-metallicity star-forming environments. The combination of the observational constraints for all three may be best explained by 
deflagrations of near $M_{\rm Ch}$ hybrid carbon-oxygen-neon white dwarfs which have
short evolutionary pathways to formation.

\end{abstract}

\keywords{supernovae: general, supernovae: individual}


\section{Introduction} \label{sec:intro}

Supernovae of type Ia (SNe Ia) are widely accepted to be explosions resulting from thermonuclear runaway in degenerate carbon-oxygen (CO) white dwarfs (WDs) in close binary systems \citep{1960ApJ...132..565H,1984ApJ...286..644N}. 
They constitute a remarkably homogeneous subclass of explosions that follow the width-luminosity relation \citep[eg.][]{1993ApJ...413L.105P,1996AJ....112.2391H}, but the  
precise nature of the progenitor, and the details of the explosion mechanism, remain open questions \citep{2011NatCo...2..350H,2014ARA&A..52..107M}. 

SNe Iax \citep{2013ApJ...767...57F} are a peculiar subclass of Ia events, named after the prototypical Iax event SN 2002cx \citep{2003PASP..115..453L}. SNe Iax are characterized by low ejecta velocities ($\sim 2000-8000$ km s$^{-1}$) and typically low luminosities, although they span a wide range in luminosity, from $M \simeq -14$ for SN 2008ha \citep{2009AJ....138..376F,2009Natur.459..674V} to $M \simeq -19$ for SN 2008A \citep{2014ApJ...786..134M}. Unlike other thermonuclear events, late time spectra of SNe Iax do not exhibit a true nebular phase, with permitted lines of (mainly) Fe~{\sc ii} persisting well beyond a year past maximum light \citep[][and references therein]{2017hsn..book..375J}. While a majority of SNe Iax show a positive correlation between peak luminosity and expansion velocity \citep{2010ApJ...720..704M}, notable outliers like SN 2009ku \citep{2011ApJ...731L..11N} and SN 2014ck \citep{2016MNRAS.459.1018T} are known to exist. 

At the extreme faint end of the objects that are broadly classified as SNe Iax are the 
low energy explosions SN\,2008ha and SN\,2010ae 
\citep{2009AJ....138..376F,2009Natur.459..674V,2010ApJ...708L..61F,2014A&A...561A.146S}.
With absolute peak magnitudes of $M_V = -14.2$ and $-13.8 \gtrsim M_V \gtrsim -15.3$ respectively, their physical nature has been disputed. 
A core collapse scenario has been suggested for SN 2008ha \citep{2009Natur.459..674V}. However this would require these two members of the Iax subclass to be distinct (along with possibly others), whereas the evidence seems to suggest a kinship with the more luminous members of the Iax family \citep{2017hsn..book..375J}.
A progenitor scenario involving a weak deflagration of a white dwarf leaving a bound remnant behind has been proposed \citep[e.g.][]{2012ApJ...761L..23J,2013MNRAS.429.2287K,2014MNRAS.438.1762F} to explain the peculiarities of SNe Iax.

This Letter reports the discovery of SN 2019gsc (ATLAS19mbg) by the ATLAS survey \citep[][]{2018PASP..130f4505T}, 
and results of detailed follow-up with Pan-STARRS1, the Nordic Optical Telescope (NOT) and Gemini-North. We note the independent study on SN 2019gsc presented by \citet{2020arXiv200200393T}, and briefly compare our results with theirs in section~\ref{sec:discussion}.

\section{Discovery and follow-up}\label{sec:disc}

SN 2019gsc (ATLAS19mbg) was discovered by ATLAS on 2019-06-02.36 UT (MJD 58636.36) in the cyan filter, at a magnitude of $c = 19.7 \pm 0.2$ \citep{2019TNSAN..23....1S}. ATLAS is a twin 0.5m telescope system on the islands of Haleakala and Mauna Loa. With a field of view $\sim$29 deg$^2$, each telescope surveys the sky robotically above declination $-40^{\circ}$ with a cadence of 2 days \citep{2018PASP..130f4505T}. The images are obtained in the 2 filters cyan and orange, that are roughly equivalent to SDSS $g+r$ and $r+i$, respectively. Using the \emph{Lasair} broker \citep{2019RNAAS...3a..26S}, we note a prior detection by the Zwicky Transient Facility \citep[ZTF;][]{2019PASP..131a8002B} on 2019-06-01.23 UT (MJD 58635.23), 
at $g = 19.85 \pm 0.17$. The transient was not detected in ZTF images on MJD 58632.28 and 58628.30, to a limiting magnitude of 20.46 and 20.50, respectively, in the ZTF $r$ filter. The transient was subsequently classified as a type Iax event by \citet{2019TNSAN..25....1L}, who noted similarity of its spectral features with the Iax SN 2010ae. The host galaxy SBS 1436+529A has a redshift of $z=0.0113$ or a heliocentric recessional velocity of 3388 km s$^{-1}$ \citep{2015A&A...578A.110A}. Correcting the velocity for the effects of Virgo infall, Great Attractor and Shapley supercluster, we adopt a distance modulus of $\mu = 33.60$ mag (assuming $H_0 = 73$ km s$^{-1}$ Mpc$^{-1}$). 

Follow-up photometry (Table~\ref{tab:magtable}) was obtained using the 1.8m \PS\ telescope \citep{Chambers2016} on Haleakala  equipped with a 1.4 Gigapixel camera (GPC1, 0.26 arcsec pixel$^{-1}$). 
Images were obtained in the \grizy\ filters \citep{2012ApJ...750...99T} and processed with the Image Processing Pipeline (IPP) described in \citet{Magnier2016data}. 
Due to SN 2019gsc being superposed to its host galaxy, image subtraction was essential for all epochs of photometry.  The Pan-STARRS1 Science Consortium \citep[PS1SC,][]{Chambers2016}
3$\pi$ survey data were used as templates, point-spread-function fitting photometry was carried out 
and photometric calibration was done against PS1 reference stars in the field \citep{waters2017,magnier2017b}. 

\begin{figure*} 
    \centering
    \includegraphics[width=0.7\linewidth]{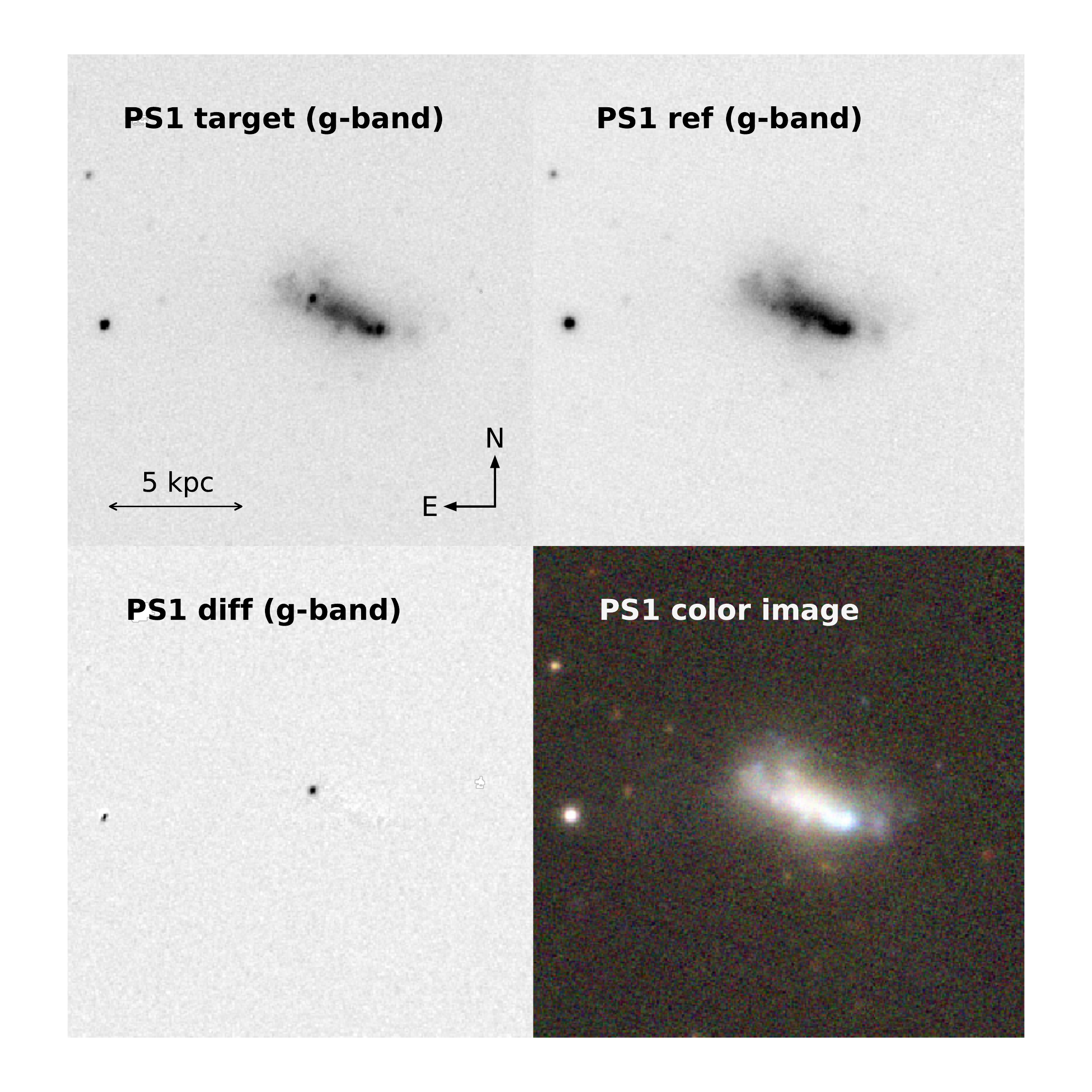}
    \includegraphics[width=0.7\linewidth]{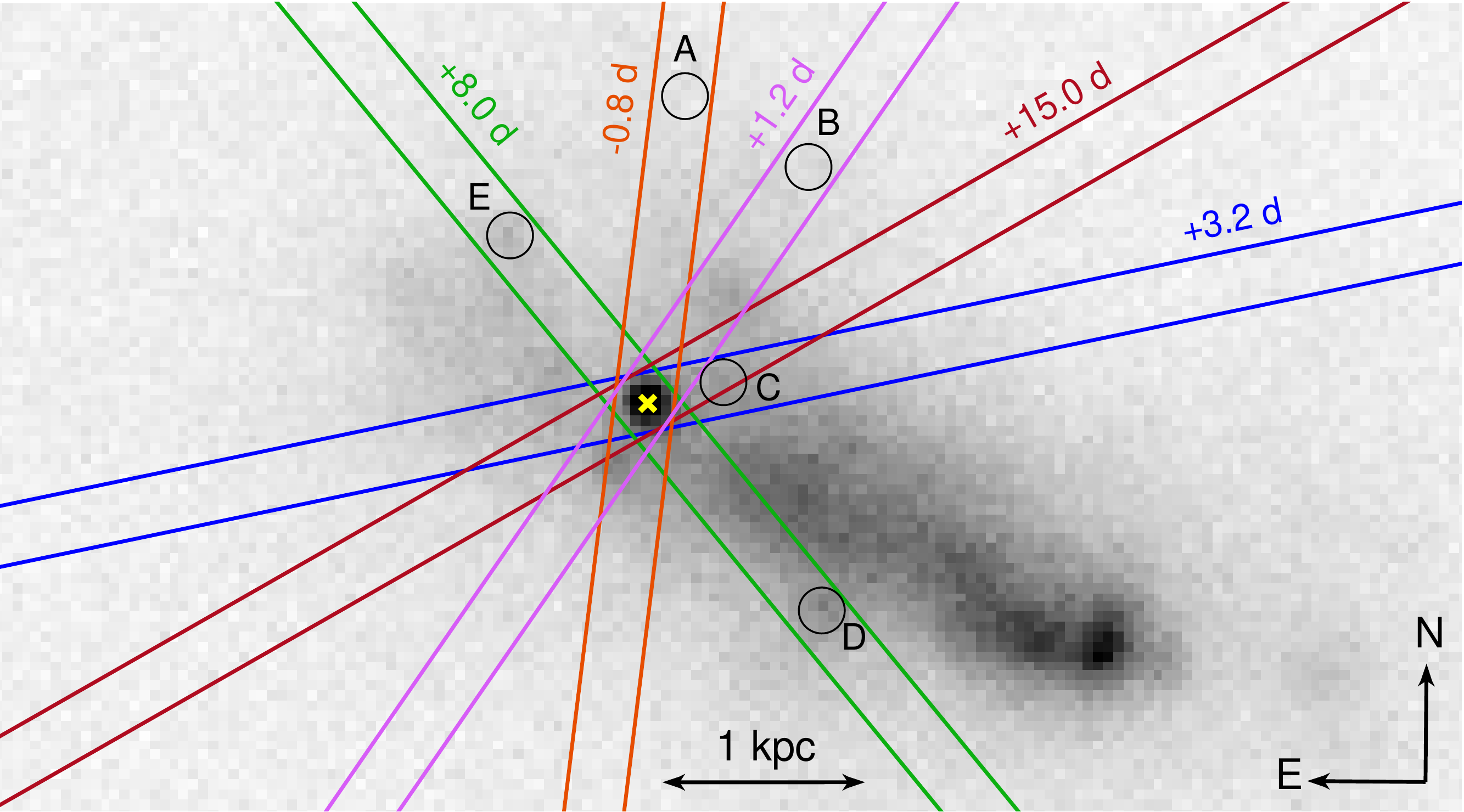}
    \caption{Top: Pan-STARRS target, reference, difference and color composite images of the field for SN 2019gsc. The $g$-band target image was acquired on MJD 58643.3 ($+4.4\,$d since \gmax). Bottom: NOT $r$-band image of SN\,2019gsc (cross) and its host galaxy, taken on MJD 58638.0 ($-0.8\,$d since \gmax). Also shown is the orientation of the slits of the NOT spectra. The phase is relative to the epoch of \gmax\ in the SN rest frame. The circles mark those regions with bright nebular emission lines used to estimate the gas metallicity (Section~\ref{subsec:metallicity}).}
    \label{fig:img}
\end{figure*}

Follow-up spectra (Table~\ref{tab:speclog}) were obtained with the Alhambra Faint Object Spectrograph and Camera (ALFOSC) on NOT during five epochs between $-0.8\,$d and $+15.0\,$d relative to \gmax, using grism 4 ($3300-9600$\,\AA) and a $1.3\arcsec$ slit yielding a resolution $R \approx 400$. The extractions, wavelength and flux calibrations were applied using custom \textsc{iraf} scripts. A  Gemini spectrum was obtained on $+25.2\,$d using the GMOS-N instrument. The GMOS spectrum, obtained using the R400 grating ($R \approx 1900$) and a $1 \arcsec$ slit, was reduced with  the Gemini \textsc{iraf} package. Synthetic photometry was computed for the spectra using the Synthetic Magnitudes from Spectra code \citep[\textsc{sms};][]{2018MNRAS.475.1046I} and the spectral fluxes were
scaled to match the multi-band \PS\ photometry. 
\PS\ images of SN 2019gsc are shown in Figure~\ref{fig:img}, along with the NOT $r$-band image and slit positions for the NOT spectra.

\begin{table*}
\centering
\caption{Summary of photometric observations of SN 2019gsc from ZTF and \PS. The phase (in days) is relative to the epoch of \gmax\ on MJD 58638.82 in the SN rest frame.}
\begin{tabular}{cccccccc}
\hline
MJD & Phase & $g$ & $r$ & $i$ & $z$ & $y$ & Instrument \\ 
\hline
58628.25 & $-$10.45 & $>20.62$          & $>20.49$         & $-$              & $-$              & $-$               & ZTF \\
58632.28 & $-$6.47  & $-$               & $>20.44$         & $-$              & $-$              & $-$               & ZTF \\
58635.23 & $-$3.55  & 19.85 $\pm$ 0.17  & 19.93 $\pm$ 0.13 & $-$              & $-$              & $-$               & ZTF \\  
58638.22 & $-$0.59  & 19.92 $\pm$ 0.18  & 19.72 $\pm$ 0.12 & $-$              & $-$              & $-$               & ZTF \\
58639.38 & +0.55    & 19.88 $\pm$ 0.05  & 19.68 $\pm$ 0.04 & 19.80 $\pm$ 0.04 & 19.87 $\pm$ 0.05 & 19.88 $\pm$ 0.19  & PS1 \\
58641.26 & +2.41    & 19.99 $\pm$ 0.09  & 19.63 $\pm$ 0.07 & 19.76 $\pm$ 0.06 & 19.74 $\pm$ 0.07 & 19.74 $\pm$ 0.21  & PS1 \\
58642.26 & +3.40    & 20.06 $\pm$ 0.09  & 19.63 $\pm$ 0.09 & 19.93 $\pm$ 0.12 & 19.82 $\pm$ 0.14 & 20.04 $\pm$ 0.33  & PS1 \\
58643.26 & +4.39    & 20.13 $\pm$ 0.04  & 19.69 $\pm$ 0.03 & 19.72 $\pm$ 0.02 & 19.78 $\pm$ 0.03 & $-$               & PS1 \\
58644.27 & +5.39    & 20.15 $\pm$ 0.06  & 19.70 $\pm$ 0.04 & 19.79 $\pm$ 0.04 & 19.74 $\pm$ 0.04 & 20.26 $\pm$ 0.17  & PS1 \\
58645.41 & +6.52    & 20.46 $\pm$ 0.18  & 19.78 $\pm$ 0.09 & 19.70 $\pm$ 0.07 & 19.69 $\pm$ 0.10 & $-$               & PS1 \\
58646.31 & +7.41    & 20.56 $\pm$ 0.12  & 19.80 $\pm$ 0.04 & 19.80 $\pm$ 0.03 & 19.81 $\pm$ 0.04 & 20.04 $\pm$ 0.12  & PS1 \\
58647.35 & +8.43    & 20.81 $\pm$ 0.13  & 19.93 $\pm$ 0.04 & 19.86 $\pm$ 0.03 & 19.88 $\pm$ 0.04 & 20.13 $\pm$ 0.13  & PS1 \\
58648.36 & +9.43    & 20.74 $\pm$ 0.16  & 19.92 $\pm$ 0.06 & 19.86 $\pm$ 0.04 & 19.83 $\pm$ 0.05 & 20.24 $\pm$ 0.18  & PS1 \\
58649.34 & +10.40   & 20.99 $\pm$ 0.28  & 19.94 $\pm$ 0.08 & 19.93 $\pm$ 0.05 & 19.89 $\pm$ 0.05 & 20.19 $\pm$ 0.16  & PS1 \\
58650.27 & +11.32   & 21.46 $\pm$ 0.24  & 20.15 $\pm$ 0.06 & 20.09 $\pm$ 0.04 & 20.01 $\pm$ 0.04 & 20.11 $\pm$ 0.10  & PS1 \\
58651.30 & +12.34   & 20.91 $\pm$ 0.25  & 20.16 $\pm$ 0.09 & 20.11 $\pm$ 0.07 & 20.04 $\pm$ 0.08 & 20.15 $\pm$ 0.22  & PS1 \\
58652.29 & +13.32   & $-$               & 20.43 $\pm$ 0.23 & 20.26 $\pm$ 0.14 & 20.04 $\pm$ 0.09 & $-$               & PS1 \\
58656.27 & +17.26   & 21.55 $\pm$ 0.25  & 20.58 $\pm$ 0.07 & 20.38 $\pm$ 0.08 & 20.23 $\pm$ 0.14 & 20.29 $\pm$ 0.30  & PS1 \\
58662.28 & +23.20   & 21.64 $\pm$ 0.13  & 20.83 $\pm$ 0.09 & 20.74 $\pm$ 0.09 & 20.45 $\pm$ 0.08 & 20.50 $\pm$ 0.25  & PS1 \\
58666.28 & +27.15   & 22.24 $\pm$ 0.27  & 21.11 $\pm$ 0.07 & 21.01 $\pm$ 0.06 & 20.64 $\pm$ 0.07 & $-$               & PS1 \\
58670.28 & +31.11   & $-$               & 21.60 $\pm$ 0.25 & $-$              & 20.70 $\pm$ 0.11 & $-$               & PS1 \\
58674.28 & +35.06   & $-$               & 21.56 $\pm$ 0.27 & 21.10 $\pm$ 0.15 & 20.79 $\pm$ 0.12 & $-$               & PS1 \\
58688.28 & +48.91   & $-$               & 22.17 $\pm$ 0.20 & 21.95 $\pm$ 0.15 & 21.22 $\pm$ 0.11 & $-$               & PS1 \\

\hline
\end{tabular}
\label{tab:magtable}
\end{table*}

\begin{table}
    \centering
    \caption{Log of spectroscopic observations for SN 2019gsc. The phase is relative to the epoch of \gmax\ in the SN rest frame.}
    \resizebox{\columnwidth}{!}{
    \begin{tabular}{ccccc}
    \hline
    Date         & MJD      & Phase   & Instrument    & Exposure        \\
    (yyyy/mm/dd) &          & (days)  &               & (s)             \\ \hline
    2019/06/03   & 58637.97 & $-0.84$ & ALFOSC/NOT    & $4 \times 600$  \\
    2019/06/06   & 58640.00 & +1.17   & ALFOSC/NOT    & $4 \times 600$  \\
    2019/06/08   & 58642.09 & +3.23   & ALFOSC/NOT    & $4 \times 600$  \\
    2019/06/12   & 58646.88 & +7.97   & ALFOSC/NOT    & $4 \times 900$  \\
    2019/06/20   & 58654.02 & +15.03  & ALFOSC/NOT    & $4 \times 900$  \\
    2019/06/30   & 58664.28 & +25.18  & GMOS-N/Gemini & $2 \times 1200$ \\
    \hline
    \end{tabular}
    }
    \label{tab:speclog}
\end{table}

\section{Light Curves and Luminosity}{\label{sec:lc}}

\subsection{Line of sight reddening}{\label{subsec:red}}

In general, colors of SNe Ia around maximum light and the color evolution after maximum can be used to estimate the host galaxy extinction \citep[eg.][]{2005ApJ...620L..87W,2010AJ....139..120F,2014ApJ...789...32B}. However, these empirical relations are unreliable for SNe Iax due to a large scatter in their intrinsic colors \citep{2013ApJ...767...57F}. 
From the NASA/IPAC Extragalactic Database (NED), the Galactic extinction along the line of sight for the host galaxy SBS 1436+529A is $A_V = 0.026$ mag \citep{2011ApJ...737..103S}, for a standard reddening law with $R_V = 3.1$. 
We do not detect any obvious Na~{\sc i} absorption in the optical spectra at the host galaxy redshift, suggesting a low host reddening for SN 2019gsc. Thus, we assume a total extinction of $E(B-V)_{\rm tot} = E(B-V)_{\rm MW} = 0.01$ mag for SN 2019gsc. For SN 2008ha, we consider $E(B-V)_{\rm tot} = E(B-B)_{\rm MW} = 0.08$ mag, following \citet{2009AJ....138..376F}. In the case of SN 2010ae, prominent Na~{\sc i} absorption in the spectra at the host galaxy redshift indicated a significant, albeit highly uncertain host galaxy extinction of $E(B-V)_{\rm host} = 0.50 \pm 0.42$ mag, with $E(B-V)_{\rm tot} = 0.62 \pm 0.42$ mag \citep{2014A&A...561A.146S}. Correcting the magnitudes for the extinction values stated above, we note that the $(g-r)$ color evolution of SNe 2019gsc and 2008ha is very similar. Adopting a moderate extinction value of $E(B-V)_{\rm tot} \approx 0.30$ mag, we see that the $(g-r)$ color evolution of SN 2010ae matches that of the other two events more closely (Figure~\ref{fig:lc_comp}). Thus, we adopt $E(B-V)_{\rm tot} = 0.30$ mag for SN 2010ae in the subsequent analysis.

\subsection{Pan-STARRS1 light curves}

\begin{figure*}
    \centering
    \includegraphics[width=0.9\linewidth]{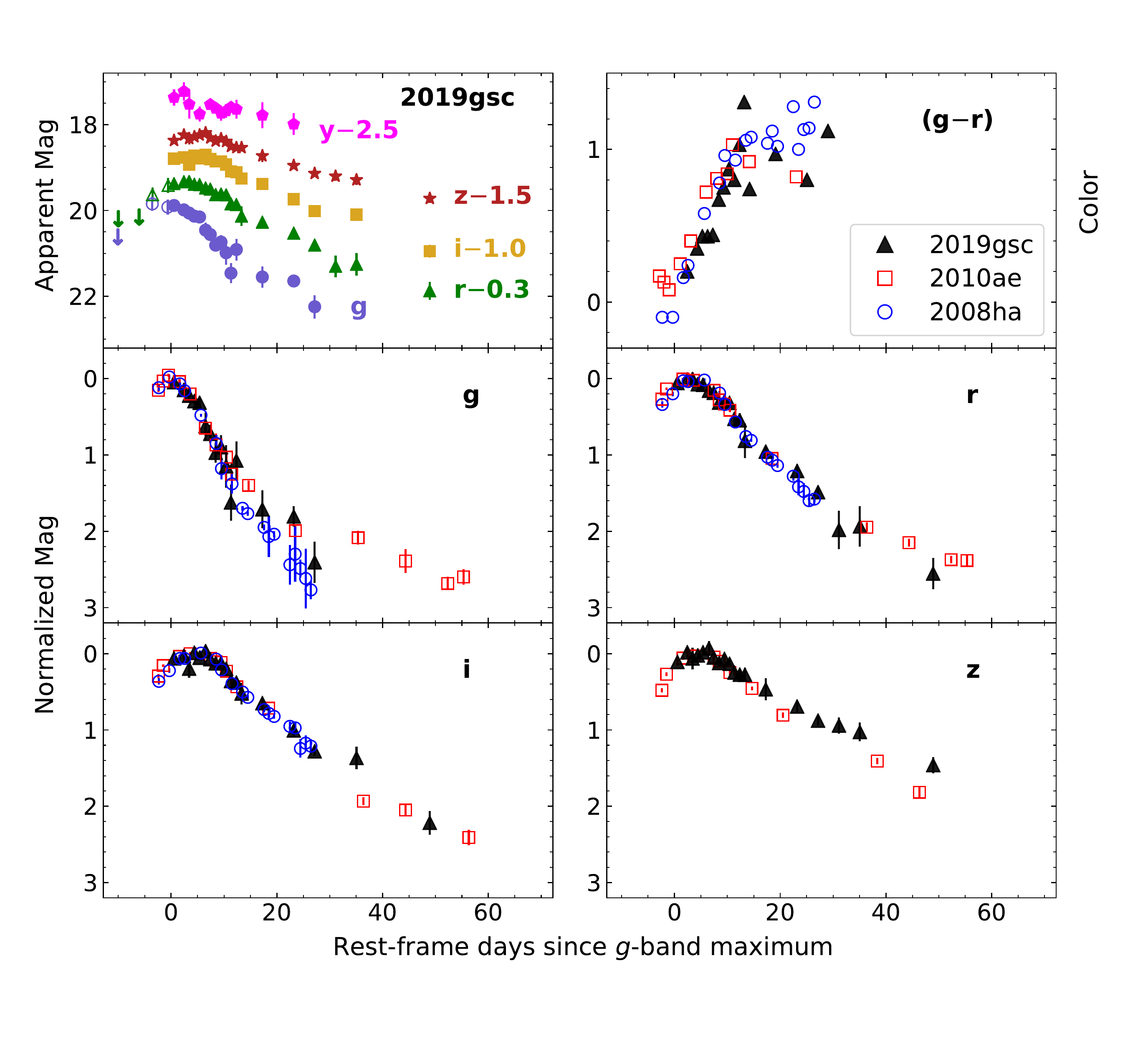}
    \caption{Pan-STARRS $grizy$ light curves of SN 2019gsc (upper left panel), plotted along with ZTF $g$ and ZTF $r$ filter magnitudes (open symbols) and ZTF upper limits (downward-pointing arrows). Also plotted is the $(g-r)$ color evolution of SNe 2019gsc, 2010ae and 2008ha (upper right panel), corrected for extinction as described in Section~\ref{subsec:red}.
    Panels 3-6 show the light curve comparison in $griz$ bands between SN 2019gsc and the faint Iax events 2008ha and 2010ae \citep{2014A&A...561A.146S}.}
    \label{fig:lc_comp}
\end{figure*}

The \PS\ (PS1) \grizy\ light curves of SN 2019gsc are shown in Figure~\ref{fig:lc_comp} (top left panel). 
The peak magnitudes and epochs in different bands, and the post-maximum decline rates were estimated by fitting lower order polynomials to the light curves. SN 2019gsc peaked in the $g$-band on MJD $58638.8 \pm 0.4$, at an apparent AB magnitude of $ m_g = 19.88 \pm 0.10$. The peak in the redder $riz$ bands occurred at later epochs, at +2.3, +3.9 and +4.4 days since \gmax, respectively.
The light curves of SNe Iax are known to be quite heterogeneous 
\citep{2016A&A...589A..89M}. SNe Iax typically show a fast decline in their light curves relative to normal SNe Ia, with decline rates comparable to the transitional and subluminous SN 1991bg-like population \citep{2015A&A...573A...2S}. 

The $griz$ light curves of SN 2019gsc are compared with those of SNe 2010ae and 2008ha \citep{2014A&A...561A.146S}, two of the faintest known Iax events, in Figure~\ref{fig:lc_comp} (panels 3--6). The light curves were normalized and shifted along the time axis to correspond to the epoch of their respective $g$-band maxima. 

\subsection{Bolometric Light Curve and Explosion Parameters}

The peak absolute magnitudes of SN 2019gsc corrected for extinction (section~\ref{subsec:red}) and assuming a distance modulus of $33.60$ mag (section~\ref{sec:disc}), $M_g^{\rm peak} = -13.75 \pm 0.23$ and $M_r^{\rm peak} = -13.97 \pm 0.16$, making it one of the faintest SNe Iax ever observed, if not the faintest. 

\begin{figure*}
    \hspace{-0.5cm}
    \centering
    \includegraphics[width=0.7\linewidth]{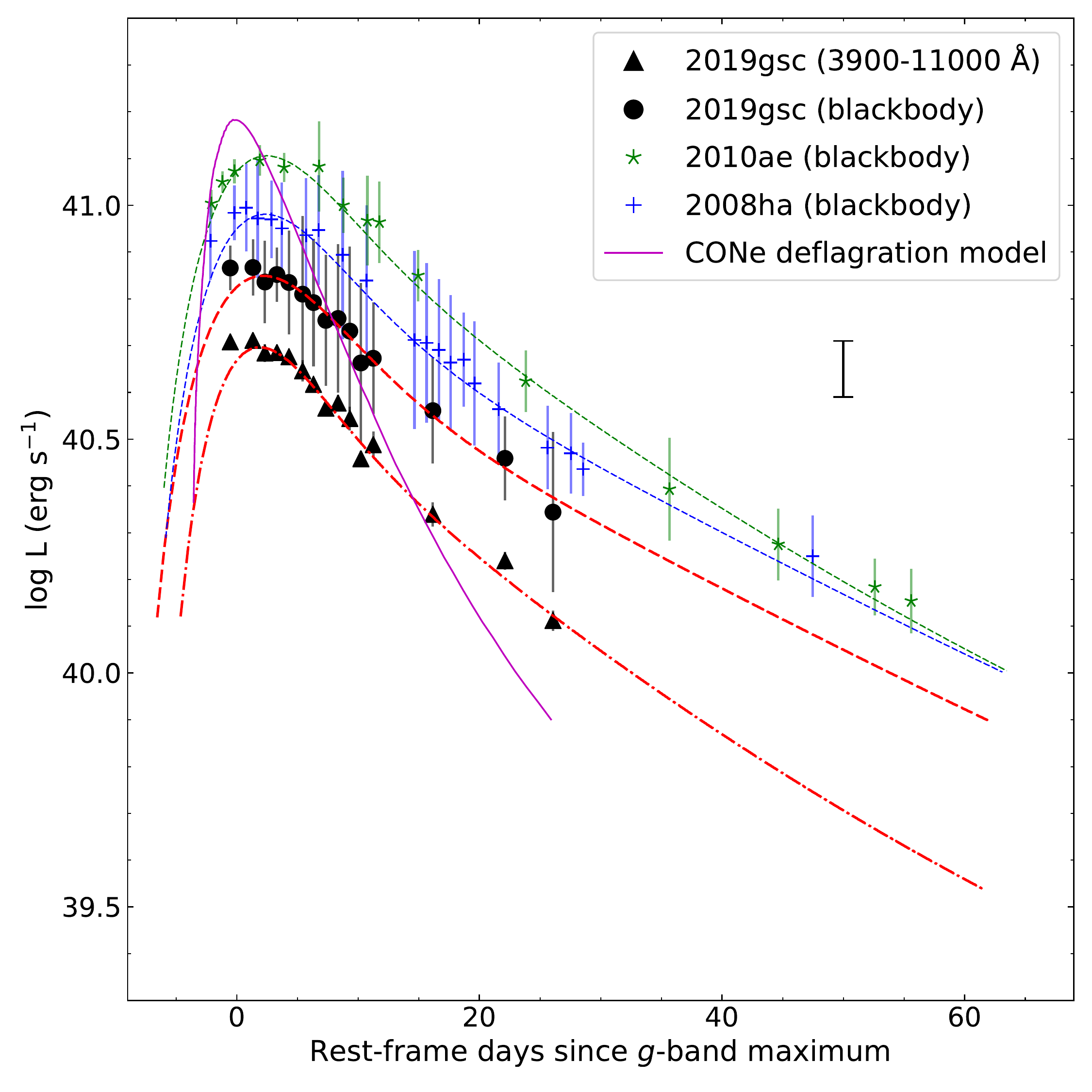}
    \caption{Quasi-bolometric ($3900-11000 \, {\mathrm \AA}$) and blackbody bolometric ($1000-25000 \, {\mathrm \AA}$) light curves of SN 2019gsc, along with the blackbody bolometric light curves of SNe 2010ae and 2008ha for comparison. The bolometric light curves were calculated using a blackbody fit to the SEDs at individual epochs, as described in the text. The dashed lines indicate the best-fiting Arnett-Valenti models. The solid line indicates the angle averaged synthetic bolometric light curve for the hybrid CONe WD deflagration model \citep{2015MNRAS.450.3045K}. The error bar below the legend represents the systematic uncertainty on the luminosity for $\pm 0.15$ mag uncertainty on the distance modulus. 
    }
    \label{fig:bolplot}
\end{figure*}

The bolometric light curve of SN 2019gsc was calculated from the multi-band PS1 photometry using the \texttt{SuperBol} code \citep{2018RNAAS...2d.230N}. 
To account for missing flux in the UV and IR bands, the code also performs a blackbody fit to the spectral energy distribution (SED) for each epoch, where a suppression factor for the UV flux can be supplied to account for line-blanketing effects.
The quasi-bolometric light curve of SN 2019gsc ($3900-11000 \, {\mathrm \AA}$), along with the full blackbody bolometric fit ($1000-25000 \, {\mathrm \AA}$), is shown in Figure~\ref{fig:bolplot}. 
We recalculated the blackbody bolometric light curves of SNe 2008ha and 2010ae from the data in \citet{2009AJ....138..376F}, \citet{2014A&A...561A.146S}, and \citet{2014A&A...561A.146S} for consistency. 
Distance moduli of $\mu = 31.64$ mag for SN 2008ha \citep{2009AJ....138..376F} and $\mu = 30.58$ mag for SN 2010ae \citep{2014A&A...561A.146S} were assumed, along with extinction corrections to the broadband as discussed in Section~\ref{subsec:red}. 
SN 2019gsc is an exceptionally low luminosity event, with a peak bolometric luminosity of $L_{\rm peak} \approx 5.1^{+0.8}_{-0.7} \times 10^{40}$ erg s$^{-1}$ for the quasi-bolometric light curve, and $L_{\rm peak} \approx 7.4^{+1.1} _{-1.0} \times 10^{40}$ erg s$^{-1}$ for the full blackbody bolometric light curve (Figure~\ref{fig:bolplot}).

The bolometric light curves were fit with an Arnett model \citep{1982ApJ...253..785A}, as formulated by \citet{2008MNRAS.383.1485V}, to estimate the explosion parameters such as $^{56}$Ni mass ($M_{\rm Ni}$), ejecta mass ($M_{\rm ej}$), and kinetic energy ($E_{51}$, expressed in units of $10^{51}$ erg). The model  assumes  homologous expansion, spherical symmetry, optically thick ejecta, and no mixing for $^{56}$Ni. The free parameters were $M_{\rm Ni}$, $M_{\rm ej}$, and the rise time $t_{\rm rise}$ of the bolometric light curve, while we fixed the optical opacity $\kappa_{\rm opt} = 0.1$ cm$^{2}$ g$^{-1}$.
The photospheric velocity was fixed at $v_{\rm ph} = 3500$ km s$^{-1}$
(see Section~\ref{sec:spec}). Fitting the quasi-bolometric ($3900-11000 \, {\mathrm \AA}$) as well as the blackbody bolometric ($1000-25000 \, {\mathrm \AA}$) light curve yields $M_{\rm Ni} \sim (1.4-2.4) \times 10^{-3}$ \msol, $M_{\rm ej} \sim 0.13-0.22$ \msol\ and $E_{51} \sim 0.01-0.02$. The fit favours a low rise time of $t_{\rm rise} \approx 10$ days, similar to that inferred for SNe 2008ha \citep{2009AJ....138..376F} and 2010ae \citep{2016A&A...589A..89M}. 

We note that the peak quasi-bolometric flux is $\sim 70 \%$ of the peak blackbody bolometric flux for SN 2019gsc, indicating a significant contribution from the UV and NIR. In contrast, the UV and NIR fraction was seen to be only $\sim 10\%$
for the transitional Ia SN 2011iv \citep{2018A&A...611A..58G}.
This could imply either a higher intrinsic UV contribution for 2019gsc, or that the UV suppression factor fed to \texttt{SuperBol} was too low, or both.

A similar analysis on the bolometric light curves of SNe 2008ha and 2010ae yields $M_{\rm Ni} \sim 3 \times 10^{-3}$ \msol, $M_{\rm ej} \sim 0.1-0.2$ \msol\ and $E_{51} \sim 0.01$; $M_{\rm Ni} = (3-4) \times 10^{-3}$ \msol, $M_{\rm ej} \sim 0.2-0.3$ \msol, $E_{51} \sim 0.03-0.05$, respectively.

\section{Spectral modelling }{\label{sec:spec}}

\begin{figure*}
    \centering
    \includegraphics[width=0.9\linewidth]{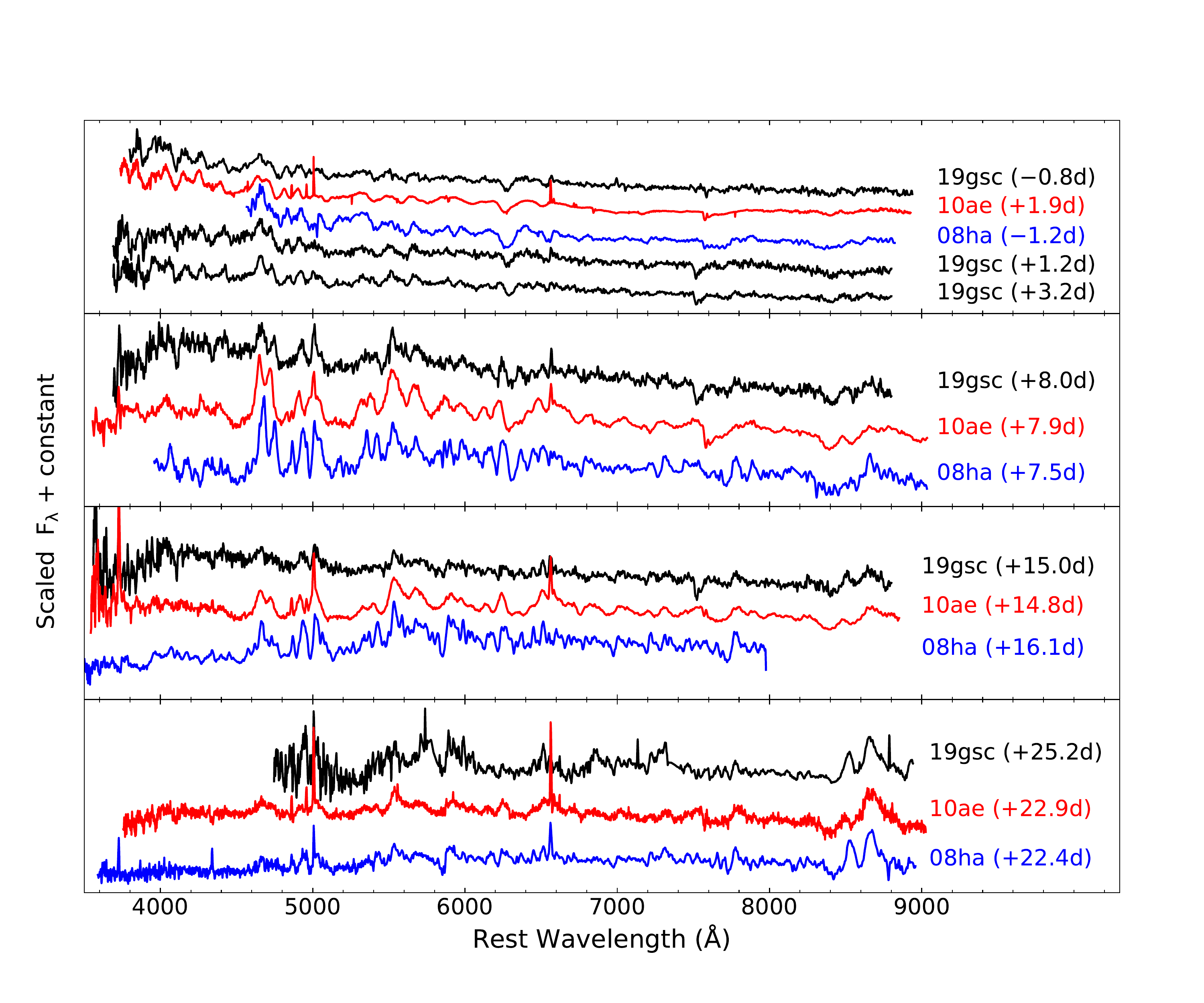}
    \caption{Spectral evolution of SN 2019gsc from $-0.8\,$d to $+25.2\,$d, compared to SNe 2010ae \citep{2014A&A...561A.146S} and 2008ha \citep{2009AJ....138..376F,2009Natur.459..674V} at similar epochs. The spectra were corrected for redshift and line of sight extinction as described in the text.} 
    \label{fig:spec_comp}
\end{figure*}

The spectra of SN 2019gsc (Figure~\ref{fig:spec_comp})  resemble the faint SNe Iax 2008ha \citep{2009AJ....138..376F,2009Natur.459..674V} and 2010ae \citep{2014A&A...561A.146S}. The spectra of SNe 2008ha and 2010ae were downloaded from the Weizmann interactive supernova data repository \citep[WISeREP\footnote{https://wiserep.weizmann.ac.il};][]{2012PASP..124..668Y}.
The earliest spectrum ($-0.8\,$d) exhibits a blue continuum with $T_{\rm bb} \approx 11000\,$K and Si~{\sc ii} $\lambda 6355$ velocity of $\sim 3800$ km s$^{-1}$, decreasing to $T_{\rm bb} \approx 8000\,$K and velocity of $\sim 1600$ km s$^{-1}$ at $+15\,$d.

We used the 1D radiative transfer code \tardis\ \citep{2014MNRAS.440..387K,kerzendorf_wolfgang_2018_1292315} 
to produce synthetic spectra to compare with the
observed data.  
\tardis\ has been previously used to model spectral sequences of SNe Iax in order to investigate the chemical structure \citep{2016A&A...589A..89M,2017A&A...601A..62M,2017MNRAS.471.4865B,2018MNRAS.480.3609B} and presence of helium in the ejecta \citep{2019A&A...622A.102M}.
\tardis\ takes a model for the ejecta with arbitrary density and abundance profiles as input, along with a luminosity and time since explosion. 
A sharp photosphere emitting a blackbody continuum is assumed, and the region above the photosphere is divided into multiple, spherically symmetric cells. The photospheric approximation assumed in \tardis\ limits its applicability to the early phase. The synthetic spectrum is calculated by iterative computation of the ionization and excitation states in each of the cells \citep{2014MNRAS.440..387K}.

For each spectral epoch of SN 2019gsc in our fit, only the time-dependent parameters (luminosity, time since explosion, inner boundary of the computation volume, and mass fractions of radioactive isotopes) were varied. The luminosity supplied to the input file was calculated by interpolating the quasi-bolometric light curve at the desired epochs. 
An exponential density profile was adopted for the SN ejecta, where the density profile is a function of velocity and time since explosion, expressed as 
$$\rho (v, t_{\rm exp}) = \rho_0 \, (t_0 / t_{\rm exp})^3 \, e^{-v/v_0}.$$

Here, we set $t_0 = 2$ days, $\rho_0 = 2 \times 10^{-11}$ g cm$^{-3}$ and $v_0 = 3000$ \kms. The epoch of explosion, $t_{\rm exp}$, was assumed to be MJD $\sim 58630.5$, roughly the mean of the two deep ZTF non-detections prior to discovery. For the abundance, we adopt the simplest case of a uniform abundance profile in each spherical cell. The outer velocity boundary of the models was set to $6000$ \kms. The model is composed primarily of carbon and oxygen, together constituting $\sim 96\%$ of the total mass. 

\begin{figure*}
    \centering
    \includegraphics[width=\linewidth]{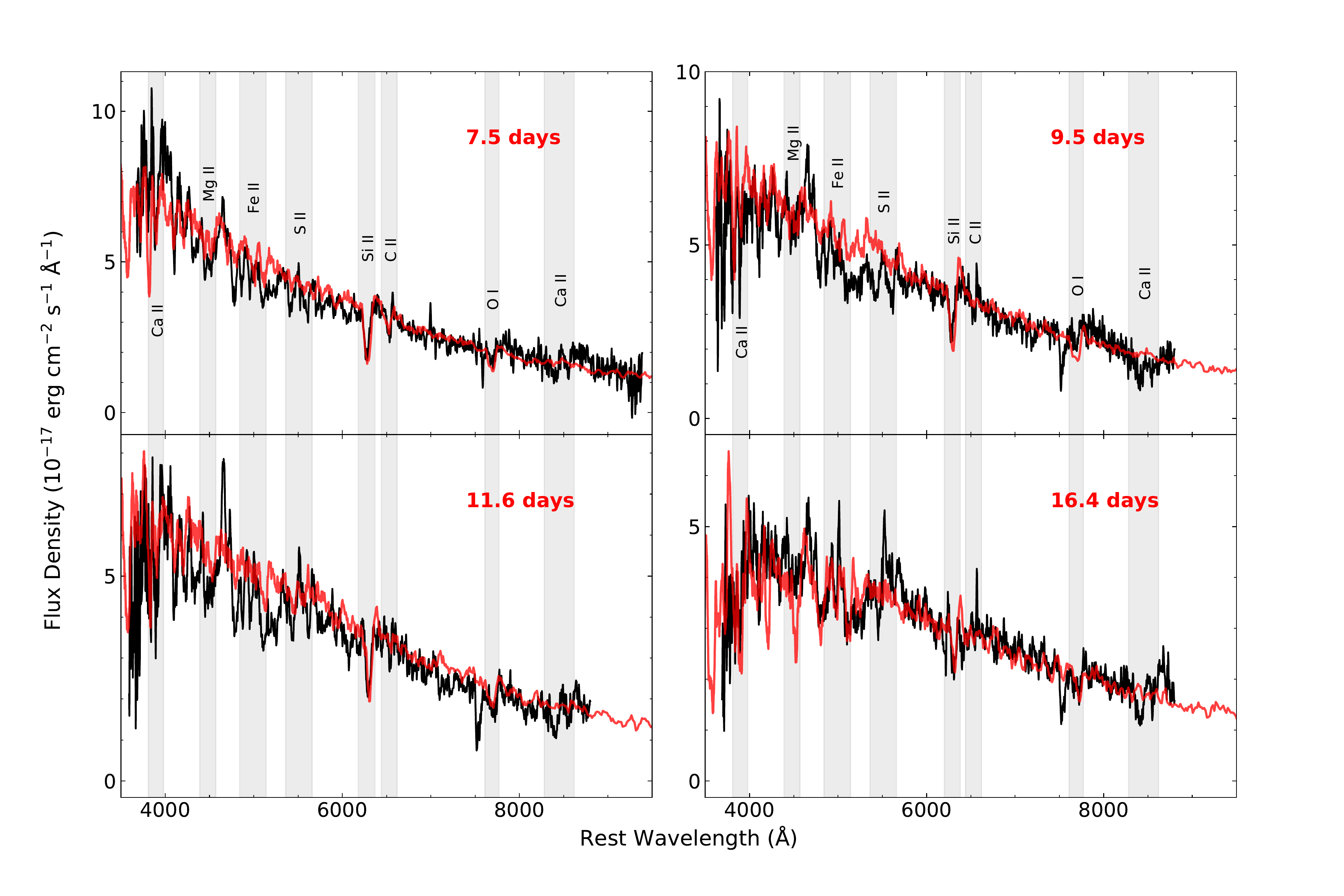}
    \caption{NOT spectra (in black) of SN 2019gsc in the photospheric phase ($7.5$ days to $16.4$ days past explosion), plotted along with the synthetic spectra (in red) generated using \tardis\ at the corresponding epoch. Shaded regions indicate prominent features in the spectra.
    The epochs correspond to the spectra in Figure~\ref{fig:spec_comp} at $-0.8$, $+1.2$, $+3.2$ and $+8.0$ days since \gmax. 
    }
    \label{fig:tardisplot}
\end{figure*}

The 4 early spectra of SN 2019gsc modeled with \tardis\ ($7.5$ to $16.4$ days past explosion) are shown in Figure~\ref{fig:tardisplot}, along with the model synthetic spectra. The continuum is well reproduced in the synthetic spectra, along with the primary spectral features of IMEs such as O~{\sc i}, Si~{\sc ii} and S~{\sc ii}. However, Fe features around $5000$\,\AA{} are not fit well, especially in the spectrum observed 9.5 days after explosion. Ca~{\sc ii} features are also identified in the spectra (Figure~\ref{fig:tardisplot}). In addition, introducing small amounts of Cr, Ti and Sc qualitatively improves the fit in the $4000-5000$~\AA\ region.

The best-fit parameters and mass fractions of various elements in the models are listed in Table~\ref{tab:composition}. In addition to the tabulated chemical elements, small amounts (${\rm X} \leq 10^{-5}$) of Na, Ca, Cr, Sc and Ti were also used.

The total mass in our \tardis\ model, summed over all the spherical shells in the computation volume amounts to $\approx 0.01$ \msol, a factor of 10 lower than the ejecta mass estimate from the bolometric light curve. This is consistent, since a significant amount of mass is expected to be below the effective photosphere delineating the optically thick and thin regions of the ejecta. The upper limit on the mass fraction of $^{56}$Ni from the model is $\lesssim 1\%$, or $M_{\rm Ni} \lesssim 10^{-4}$ \msol, consistent with the $^{56}$Ni mass estimate from the bolometric light curve. 

\begin{table*}[]
    \centering
    \caption{Best-fit parameters and mass fractions of different chemical elements in the \tardis\ models for SN 2019gsc. $v_{\rm inner}$ denotes the inner boundary of the computation volume and $t_{\rm exp}$ is the time since explosion. The emergent luminosity ($L$) was fixed by interpolating the bolometric light curve at the relevant epochs.}
    \begin{tabular}{ccccccccccc}
    \hline
    $t_{\rm exp}$ & $L$ & $v_{\rm inner}$ & $X$(C) & $X$(O) & $X$(Mg) & $X$(Si) & $X$(S) & $X$(Fe) & $X$(Co) & $X$(Ni) \\
    (day) & ($\log L_{\odot}$) & (km s$^{-1}$) & & & & &  & & & \\ \hline
    7.5 & 7.32 & 2600 &  0.75 & 0.21 & 0.01 & 0.02 & 0.005 & $2\times 10^{-4}$ & $2.8\times 10^{-4}$ & $2.1\times 10^{-4}$ \\
    9.5 & 7.35 & 2300 &  0.75 & 0.21 & 0.01 & 0.02 & 0.005 & $2\times 10^{-4}$ & $3.1\times 10^{-4}$ & $1.7\times 10^{-4}$ \\
    11.6 & 7.34 & 2000 & 0.75 & 0.21 & 0.01 & 0.02 & 0.005 & $2\times 10^{-4}$ & $3.4\times 10^{-4}$ & $1.4\times 10^{-4}$ \\
    16.4 & 7.23 & 1600 & 0.75 & 0.21 & 0.01 & 0.02 & 0.005 & $2\times 10^{-4}$ & $3.9\times 10^{-4}$ & $8\times 10^{-5}$ \\
    \hline
    \end{tabular}
    \label{tab:composition}
\end{table*}

\subsection{Host metallicity}
\label{subsec:metallicity}

While SNe Iax mostly do not appear to correlate with low metallicity environments 
\citep{2017A&A...601A..62M}, both of the very low luminosity events (SN 2008ha  and SN 2010ae) were associated with low metallicity host galaxies. \citet{2009AJ....138..376F} estimated an oxygen abundance of $12 + \mathrm{log(O/H)} = 8.16 \pm 0.15$ dex for SN 2008ha and 
from the N2 and O3N2 indicators of \citet{2004MNRAS.348L..59P} while 
\citet{2014A&A...561A.146S} estimated 
$12 + \mathrm{log(O/H)} = 8.40 \pm 0.12$ dex (N2) and $8.34 \pm 0.14$ dex (O3N2). 

In order to estimate the host metallicity for SN 2019gsc, we used the NOT spectra obtained at different dates. 
All spectra had different slit orientations, as they were obtained at parallactic angle but at different hour angles (Fig.~\ref{fig:img}). 
Regions  A--E  were dominated by strong nebular emission lines.
The [N~{\sc ii}]  lines are not significantly detected in regions A, B and C, leading to upper limits of 
$12 + \mathrm{log(O/H)} < 8.10$ (8.07) dex,
$12 + \mathrm{log(O/H)} < 8.18$ (8.13) dex; and $12 + \mathrm{log(O/H)} < 8.35$ (8.18) dex, calibrated on the N2 (O3N2) scales, respectively.
In regions D and E, [N~{\sc ii}] is marginally detected, allowing us to estimate a metallicity of $12 + \mathrm{log(O/H)} = 8.10 \pm 0.06$ dex (N2), $12 + \mathrm{log(O/H)} = 8.08 \pm 0.06$ dex (O3N2) for region E; and $12 + \mathrm{log(O/H)} = 7.81 \pm 0.19$ dex (N2), $12 + \mathrm{log(O/H)} = 7.86 \pm 0.08$ dex (O3N2) for region D. 
These values suggest a very metal-poor environment (15\% solar) for SN 2019gsc,
among the most metal-poor environments for the Iax sample \citep{2018MNRAS.473.1359L}.

In addition, the extracted spectra at regions A, B, D and E also show very strong H$\alpha$ and 
[O~{\sc iii}] $\lambda 5007$ emission lines (equivalent widths of $\sim 100-200$~\AA), suggesting that SN 2019gsc is in proximity to regions undergoing a starburst.

\section{Discussion and Conclusions}\label{sec:discussion}

SN 2019gsc is probably the least luminous supernova-like transient discovered, with a peak luminosity $M_g \approx -13.8 \pm 0.2$ mag. 
It shows strong photometric and spectroscopic similarities to 
SNe 2008ha and 2010ae, which had peak luminosities of $M_g = -14.01 \pm 0.14$, $-13.54 \gtrsim M_g \gtrsim -15.33$ \citep{2014A&A...561A.146S}. 
The explosion parameters estimated from the bolometric light curves of all three are also similar. 
Fitting the bolometric light curve of SN 2019gsc with an energy deposition model suggests $M_{\rm Ni} \sim 2 \times 10^{-3}$ \msol, 
$M_{\rm ej} \sim 0.2$ \msol\ and kinetic energy $E_{51} \sim 0.01-0.02$. Our estimates for the luminosity, decline rates and explosion parameters for SN 2019gsc are consistent, within uncertainties, with those presented in \citet{2020arXiv200200393T}. 
Furthermore, somewhat remarkably,  all three occur in blue, star-forming host galaxies, with indications of moderately low to extremely low metallicity host environments.  These three SNe represent the 
extreme low luminosity and low energy end of the population of supernovae classed as type Iax.  However the peculiar properties that they 
all have in common now raises the question if they are physically distinct explosions from the bulk of the SN Iax population.

Pure deflagrations of WDs \citep{2004PASP..116..903B,2007PASP..119..360P} have been proposed to account for the low explosion energy and luminosity of SNe Iax in general. Hydrodynamic 3D simulations of weak deflagrations can fail to completely unbind the WD and are predicted to leave a `bound' remnant behind  \citep{2012ApJ...761L..23J,2013MNRAS.429.2287K}.
However, the extremely low luminosities of SNe 2019gsc, 2010ae and 2008ha place them below the range of ejected $^{56}$Ni masses predicted by existing sets of deflagation models for CO WDs
\citep{2012ApJ...761L..23J,2014MNRAS.438.1762F}.
One potential means to obtain an extremely low $^{56}$Ni ejected mass was proposed by 
\citet{2015MNRAS.450.3045K}.  A  deflagration simulation involving a near-$M_{\rm Ch}$ hybrid carbon-oxygen-neon (CONe) WD yielded $M_{\rm Ni}\sim 3 \times 10^{-3}$ \msol, consistent with the luminosity of SNe 2019gsc, 2010ae and 2008ha. A very low total ejecta mass resulted from these simulations
($M_{\rm ej}\sim 0.01$ \msol), which is a factor 20 lower than the mass we infer from the light curves. The comparison in Figure\,\ref{fig:bolplot} illustrates the point that the ratio of 
$M_{\rm Ni}/M_{\rm ej}$ from the \citet{2015MNRAS.450.3045K} CONe WD explosion simulation is
significantly higher than that inferred from the observed light curves.  The CO deflagrations
of \citet{2014MNRAS.438.1762F} produce higher ejecta masses, more compatible with the data
(e.g. their N3def model), but the $^{56}$Ni mass from that simulation is too high. 
A uniform composition is sufficient for satisfactory \tardis\ fits to the early spectra. This favours a mixed ejecta rather than a layered ejecta composition, since the turbulent mixing in deflagration is expected to prevent distinct 
velocity layers containing different elements 
\citep{2003Sci...299...77G}. However, more complex, layered ejecta structures are not ruled out.

Binary population synthesis studies have shown that hybrid CONe WDs with helium-burning donors have short delay times of $30-180$ Myr \citep{2014ApJ...794L..28W,2015MNRAS.450.3045K}. This is consistent with the fact that Iax events are generally associated with young stellar populations \citep{2014ApJ...792...29F,2014Natur.512...54M,2018MNRAS.473.1359L}.
In addition, low metallicity stars are expected to form higher mass WDs \citep{2013ApJ...770...88K}, implying a shorter time required to bring them to explosion.
The fact that all three of these extremely low luminosity Iax are in apparently young stellar populations suggests this channel might be promising. 

Our observations  do not yet rule out a massive star origin for these faint events. A weak core-collapse of a stripped, massive star involving fallback on to the central remnant was proposed for SN 2008ha \citep{2009Natur.459..674V,2010ApJ...719.1445M}. 

The discovery of a blue point source in pre-explosion HST images of the type Iax SN 2012Z led \citet{2014Natur.512...54M} to argue for a helium-rich donor as the binary companion to the WD, rather than a massive star.  This scenario is consistent with the detection of helium features in the spectra of potential Iax events 2004cs and 2007J \citep{2013ApJ...767...57F,2016MNRAS.461..433F}, although their association with SNe Iax has been disputed \citep{2015ApJ...799...52W}. Although we do not invoke helium in our \tardis\ model, this does not rule out its presence in the ejecta \citep[see][]{2019A&A...622A.102M}. 

Future detection and characterisation of `bound'  remnants at the sites of Iax explosions \citep{2019ApJ...872...29Z}, and unambiguous spectroscopic confirmation of the companions of Iax progenitors as helium stars, would help resolve these questions. Further theoretical simulations
of deflagrations are required to explore if the model parameters can produce the observed 
$M_{\rm Ni}/M_{\rm ej}$ ratio and better reproduce the observed light curves. The balance between
how much $^{56}$Ni is trapped in a bound object, compared to how much is ejected needs 
further investigation. In addition, the energy radiated from a `bound' object may be slower than
from a freely expanding remnant, which would help slow the LC evolution.

%

\facilities{ATLAS, Pan-STARRS, NOT, Gemini-North}





\acknowledgements
 Data are from the Asteroid Terrestrial-impact
  Last Alert System project (NASA grants NN12AR55G, 80NSSC18K0284,
  and 80NSSC18K1575), byproducts of the NEO search include images and
  catalogs from the survey area, through the contributions of the University of Hawaii
  Institute for Astronomy, the Queen's University Belfast, the Space
  Telescope Science Institute, and the South African Astronomical Observatory. Pan-STARRS is 
  supported by the NASA Grants, including No. NNX14AM74G  through the SSO NEO Observations Program.
  SJS, SS, SAS, DRY and KWS acknowledge STFC Grants ST/P000312/1, ST/N006550/1, ST/N002520/1  and ST/S006109/1. 
  SS thanks B. Barna and M. R. Magee for useful discussions.
GL and DBM acknowledge research grant 19054 from VILLUM FONDEN. CG acknowledges VILLUM FONDEN research grant 25501. This work was supported by a VILLUM FONDEN Investigator grant 16599 to JH.
KAA acknowledges DNRF132. We thank Jonatan Selsing for help with NOT spectroscopic reductions.
Nordic Optical Telescope (programs 59-008 and 59-509), operated by the NOT Scientific Association at the Observatory del Roque de los Muchachos, La Palma, Spain, of the IAC.
Gemini observatory (program GN-2019A-Q-126) is operated by the Association of Universities for Research in Astronomy, Inc., under a cooperative agreement with the NSF on behalf of the Gemini partnership: the National Science Foundation (United States), National Research Council (Canada), CONICYT (Chile), Ministerio de Ciencia, Tecnolog\'{i}a e Innovaci\'{o}n Productiva (Argentina), Minist\'{e}rio da Ci\^{e}ncia, Tecnologia e Inova\c{c}\~{a}o (Brazil), and Korea Astronomy and Space Science Institute (Republic of Korea).
We made use of \tardis\ supported by the Google Summer of Code and ESA's Summer of Code in Space.

\software{https://github.com/mnicholl/superbol}






\bibliography{references}

\end{document}